\documentclass[12pt,preprint]{aastex}

\begin{document}
\title{A Far Ultraviolet Study of the Old Nova V841 Oph}

\author{Scott G. Engle and Edward M. Sion}
\affil{{Department of Astronomy \& Astrophysics}
\affil{Villanova University}
\affil{Villanova, PA 19085}
e-mail:scott.engle@villanova.edu
       edward.sion@villanova.edu}

\begin{abstract}
 
We have carried out a synthetic spectral analysis of archival IUE spectra
of the old nova V841 Oph (Nova Oph 1848) taken 15 years apart. The spectra
reveal a rising continuum shortward of 1560\AA, a C {\sc iv} P-Cygni profile
indicative of wind outflow associated with disk accretion in one spectrum,
a deep Ly $\alpha$ profile, and strong N {\sc v} (1238\AA, 1242\AA) and O {\sc v}
(1371\AA) wind/coronal absorption lines. Numerous sharp interstellar
resonance lines are also present. A grid of high gravity atmospheres and
accretion disk models, spanning a wide range of inclinations, accretion
rates and white dwarf masses was compared to the de-reddened spectra. We
find that, for a steady state accretion disk model to account for the FUV
spectra, the accretion rate is only $\sim$3 $\times$ 10$^{-11}$
M$_{\sun}$/yr, 147 years after its outburst in 1848, with an implied
distance within $\sim$300 pc. The accretion
rate at 147 years post-outburst is smaller than expected for an old
nova.

\end{abstract}

 
\section{Introduction}

Cataclysmic variables are a diverse class of semi-detached, compact,
interacting binaries in which a white dwarf (WD) primary star accretes
matter from a gaseous disk formed by mass transfer from a Roche
lobe-filling main sequence secondary star. The classical novae and
recurrent novae are subclasses of CVs whose outbursts are due to explosive
thermonuclear burning on the surface of the White dwarf. A typical
outburst will produce a maximum increase in brightness of $\sim6-15$ mag
and release $10^{44}-10^{46}$ ergs. When taking into account the observed
frequency of classical nova outbursts and the observed CV space density,
we can infer a lower limit recurrence time for a given system to be
$\sim10^4-10^5$ yr. Thus the recurrence time is very short compared to the
lifetime of a CV so that CVs will experience many classical nova outbursts
in their lifetimes. The old novae and recurrent novae are the only CVs
that have recorded evidence of having undergone nova explosions.

Among the old novae, DI Lac (Moyer et al. 2003) and V841 Ophiuchi are the
only such systems that have optical and UV spectra resembling the UX
UMa-type nova-like variables (Warner 1995 and references therein).
Emission line widths are very narrow, indicating a low inclination, as
first noted by Kraft (1964) and later confirmed by other studies. A low
inclination could explain why these two old novae have the highest optical
to X-ray flux ratios. We note however, that their optical spectra are
quite different (Hoard et al. 2000). The optical spectra of DI Lac reveal
broad, shallow Balmer and helium absorption lines with superimposed weak
emission lines, whereas V841 Oph reveals strong single-peaked Balmer, He
{\sc i} and He {\sc ii} emission features.

The impetus for this work arises from the paucity of knowledge about  
the white dwarfs and accretion disks in old
novae. How high are their accretion rates in the
post-nova state? Is there a correlation between the time since the nova
and the accretion rate in the old nova system? How quickly is an accretion
disk established following a nova explosion? How hot is the post-nova
white dwarf and how was it affected by the explosion? These basic
questions lack definitive answers and underscore the need for advancing
our knowledge of these systems. This is especially true in the FUV range
as shown in the recent analysis of the HST STIS spectrum of DI Lac (Moyer
et al. 2003).

V841 Oph was discovered at the time of its outburst in 1848, as Nova Oph 
1848 and the early literature classified it as a fast nova. However, the 
true maximum was very likely missed and hence the correct $t_{3}$ value 
remains uncertain. This uncertainty in $t_{3}$ and its impact on the speed 
class and distance estimates of V841 Oph is discussed in section 3.1 
below. Its orbital period is 0.60144$\pm$0.00056 (Ritter and Kolb 1998). 
The observed or estimated physical and orbital parameters published for 
V841 Oph are listed in Table 1.

Warner (1995) derives absolute magnitudes of CV subtypes and assigned
M$_V$ = +3.9 as an average for old novae as a class which, if due to
accretion, implies an accretion rate of 10$^{-8}$ M$_{\sun}$/yr. Warner
found a strong correlation between emission line equivalent widths and the
inclinations of the CV systems, as well as a strong correlation between
the inclination angle and the absolute magnitude. The systems having
higher inclination angles also had lower M$_V$ values than those that had
smaller inclination angles. Moreover, the systems that were nearly
pole-on, like DI Lac and V841 Oph, were found to have the highest M$_V$
values.

\section{Far Ultraviolet Archival Observations}

We extracted the archival spectra of V841 Oph from the IUE NEWSIPS final
archive. The observing log is given in Table 2 where we have listed, by
column: SWP image number, exposure time, starting time of the exposure,
the original program ID and the continuum and background counts. The
observations were made using the large aperture with the short-wavelength
prime camera (SWP) at low dispersion with a resolution of 5\AA, covering a
wavelength range of 1170-2000\AA.

We carried out a flux calibration/correction to the IUE NEWSIPS spectra
using the Massa-Fitzpatrick correction algorithm to improve data quality
and signal-to-noise. Massa \& Fitzpatrick (2000) showed that the NEWSIPS
low-dispersion data had an absolute flux calibration that was inconsistent
with its own reference model and also subject to time-dependent systematic
effects which could, added together, amount to 10-15\%. Using our
corrected fluxes, we removed the effect of interstellar reddening by using
the IDL routine UNRED with E(B-V) = 0.50 (Verbunt 1997).

The two spectra for V841 Oph were taken just over 15 years apart. However,
they are remarkably different from each other. The first spectrum has a
better signal-to-noise and a possible P Cygni profile in C {\sc iv} 1550\AA.
The more recent spectrum seems to have no discernable wind outflow, yet
also has flux levels marginally higher than the earlier spectrum. It 
lacks a distinguishable P Cygni profile, and is generally noisier than the
earlier spectrum. The lower quality of the second spectrum is due largely
to the degradation of the SWP camera which marked the last few years of
IUE operations. 

Despite the noise level in both spectra, a few strong line features are
unambiguously identifiable. The strongest features in SWP07950 are the
broad Ly $\alpha$ absorption, the aforementioned C {\sc iv} (1550\AA)
absorption, N {\sc v} (1240\AA) absorption and Si {\sc iii} + O {\sc i}
(1300\AA) absorption are also evident. The peak to peak noise level in
SWP54455 preclude a clear identification of any line feature other than
the Ly $\alpha$ absorption, likely Si {\sc iii} + O {\sc i} (1300\AA)
absorption, and N {\sc v} (1240\AA) absorption. In both spectra, there is
a hint of Si {\sc iv} (1260\AA) absorption. However, it is somewhat puzzling
that Si {\sc ii} (1260\AA) is weak, if present at all, relative to the
prediction of a solar composition accretion disk (see below). This would
suggest the possibility that Si is underabundant relative to solar
composition. This is also suggested by the absence of Si {\sc iv} wind
absorption in the face of C {\sc iv} (1550) and N {\sc v} (1240\AA), all
three of which are typically seen together in cataclysmic variable wind
outflows. A similar underabundance of Si was seen in the analysis of the
HST STIS spectrum of DI Lac (Moyer et al. 2003).

\section{Synthetic Spectral Modeling}

The TLUSTY (Hubeny 1988) and SYNSPEC (Hubeny \& Lanz 1995) codes were used
to create model spectra of white dwarf stellar atmospheres. Solar
abundances were assumed. Wade \& Hubeny's (1998) optically thick disk
model grid was the source of model accretion disks. We then used IUEFIT,
which is a $\chi^{2}$ minimization routine, to calculate a $\chi^{2}$
value and a scale factor for each of our models.

The scale factor S is defined in terms of the stellar radius R and distance
{\it d} by:

    F($\lambda_{obs}$) = S H($\lambda_{model}$)

where S = 4($\pi$)(R$^2$/{\it d}$^2$).

The scale factor is normalized to one kiloparsec for the distance and one
solar radius for the radius. Thus, for a photosphere the distance {\it d} is
given by:

{\it d} = 1000pc (R$_{wd}$/R$_{\sun}$)/ S$^{0.5}$.

Since the accretion disk model fluxes are normalized to a distance of
100 pc, for an accretion disk fit, the distance is given by:

{\it d} = 100/S$^{0.5}$pc.

We also carried out combination fits utilizing both the best-fitting
accretion disk model and the best-fitting photosphere model. With this
fit, we were able to obtain the relative contributions of the accretion
disk and the white dwarf.
    
The spectra were prepared for fitting by masking regions with negative
flux. For SWP54454 we masked 1190-1223\AA, along with 1787-1793\AA. For
SWP07950 we masked wavelengths $<$ 1228\AA, as well as 1925-1927\AA.

First, we carried out accretion disk-only fits to the data. We ran models
with white dwarf masses of 0.350, 0.550, 0.800, 1.030 and 1.210 solar
masses. There was also a range of log $\dot{M}$ values from -8.0 up to
-10.5 in increments of 0.5. The disk inclination angle {\it i} was kept
fixed at 18$\degr$\ in all of the fits to be consistent with the evidence
that V841 Oph is a very low inclination system. The parameters of the
best-fitting accretion disk models are summarized in Table 3. For the same
white dwarf mass of 0.8 M$_\sun$, the best-fitting disk-only model to
SWP07950 is displayed in fig. 1 while the best-fitting disk model to SWP54454
is displayed in fig. 2. For a fixed white dwarf mass of 1.0 M$_\sun$, the
best-fitting disk-only models to SWP07950 and SWP54454 are displayed in
figs. 3 and 4, respectively. The best-fitting accretion disk models to
both spectra have the same accretion rate, $\sim$3$\times$10$^{-11}$
M$_\sun$/yr, and the scale factors of these best-fits place V841 Oph
within 200 pc of the sun.

It is noteworthy that in both spectra, relative to the disk model fits,
the data show a broad ``bump'' centered at $\sim$1500\AA\ unaccounted for by the
model disk continuum.  While this ``bump'' could be due to underlying
emission giving the appearence of a curved continuum in that region, we
believe that another possibility is more likely, namely that there is an
underlying high velocity component (an accretion belt on the white dwarf
or an optically thick inner disk/boundary layer ring in which the
Keplerian broadening of the wings of smeared absorption lines produces the
``arched'' continuum in that region.) We have seen this in VW Hydri's white
dwarf spectrum during quiescence and regard it as a hallmark of such an
underlying high velocity component. If this is the case, then in V841 Oph,
the curvature of the continuum could be due to the merging of
Keplerian-broadened wings.

Next, we applied white dwarf photosphere models. We found that our
best-fitting models to the two spectra were achieved with a temperature of
18,000K-20,000K and log {\it g} = 8. For comparison with the disk fits, we
display the best-fitting photosphere model to SWP54454 in fig. 4 and the
best-fitting photosphere model to SWP07950 in fig. 5.

We also combined the best-fitting white dwarf photosphere fluxes with the 
best-fitting accretion disks fluxes, varying the disk fluxes by a small
increment between 0.1 and 10 until a best combination fit was achieved
(see Winter and Sion (2003) for the details of this technique). 
The resulting best-fitting photosphere plus disk fit to the two
spectra yielded distances of 260 pc for SWP07950, and 258 pc for
SWP54454, which are both close to the more recently published values by
Warner (1987). Combined models showed an overwhelming contribution from
the disk itself, with 94\% of the SWP07950 flux, and 94\% of the 
SWP54454 flux coming from the accretion disk.

Although we fixed the inclination of V841 Oph at 18$\degr$ and adopted
0.8 M$_{\sun}$ and 1.0 M$_{\sun}$ as the most likely white dwarf masses in our
accretion disk model, we extended our fits to encompass the range of
probable white dwarf masses and orbital inclinations derived by Diaz and
Ribeiro (2003) based upon probability calculations in their radial
velocity study. If we adopt an inclination as high as 60$\degr$, then for
a white dwarf mass of 0.8 M$_{\sun}$, we find that the best-fitting accretion
disk model has an accretion rate of 3 x 10$^{-10}$ M$_{\sun}$/yr and a
scale factor-derived distance of 283 pc. If we adopt a white dwarf mass of
0.35 M$_{\sun}$, which lies close to the lower end of the Diaz and Ribeiro
(2003) mass range, then we obtain a best-fitting disk model with an
accretion rate of 1 x 10$^{-9}$ M$_{\sun}$/yr and a model-derived distance
of 229 pc. While we believe that a white dwarf mass
this small would imply a helium-rich core, low ejection velocities and
extremely long nova recurrence timescales, recent hydrodynamic simulations
of classical novae extended to lower white dwarf masses and very low accretion
rates do indeed result in nova events (Yaron et al. 2005).
In the final analysis, all of our accretion disk model
fits imply that the distance to V841 Oph is below 300 pc.

\subsection{The Distance to V841 Oph}

The distance implied by our model fitting is considerably closer than one
typically associates with old novae. In fact, classical novae are often
found at larger distances than other CV types at least in part because
they are discovered during their bright outbursts (i.e., it is influenced
by an observational selection effect). In view of the cyclical
(hibernation) evolution scenario for CVs, it should not be surprising to
find novae at distances comparable to a ``typical'' CV (i.e., a few hundred
pc).  Given that V841 Oph is located in the direction of the Galactic
bulge where novae typically are at larger distances, the distance we have
derived for V841 Oph would be puzzling. However, it cannot be ruled out
that V841 Oph is simply a disk object that lies between the sun and the
center of the Galaxy.

The early literature classified V841 Oph as a fast nova. If the true 
maximum was missed, then the nova was brighter and the parameter $t_{3}$ 
smaller. A smaller $t_{3}$ indicates an intrinsically brighter nova. 
Payne-Gaposchkin (1957) pointed out the difficulty of reducing the 
different observations to a consistent light curve and that the true 
maximum, which she thought was near +2, was missed.  After combining this 
information with that of Pickering (1900), based on observations by 
F.W.A. Argelander, the first observation of a magnitude of +5.0 is 9 days 
later than Payne-Gaposchkin's approximate maximum. If the Pickering 
magnitudes are reliable, this suggests a $t_{3}$ value of about 12 days. 
Using the magnitude-rate of decline relationship (Downes and Duerbeck, 
2000) for a light curve of type B (Duerbeck, 1981), leads to an absolute 
magnitude of -7.5. The fast nova relationship leads to an absolute 
magnitude of -9.6. An absorption in V of 1.60 (Gilmozzi et al. 1994) from 
fitting the UV continuum, then gives corresponding distances of 380 and 
1000 pc, respectively. The lower value is closer to our determination of 
190 pc.  On the other hand, Sherrington and Jameson (1983), by fitting the 
infrared to include the contribution of the cool component, find a 
distance between 336 and 301 pc.

Since the true speed class of V841 Oph is uncertain due to different 
estimates of $t_{3}$, it is worth exploring a slow nova relationship for 
maximum magnitude (Woronsow-Weljaminow 1953; Duerbeck 1981). In this case, 
the implied distance is $>$ 370 pc for a $t_{3}$ value of 23 days. 
Duerbeck (1987) gives an apparent maximum of +4.2 and a $t_{3}$ value of 130 
days. For the same interstellar absorption, the corresponding Downes and 
Duerbeck (2000) relation gives an absolute magnitude of -6.9. The distance 
then is 790 pc. Warner (1987) gives an absolute magnitude at maximum of 
-6.5 for a maximum apparent magnitude of +4. His absorption is $A_{v}$ = 
1.25 and the distance obtained is 710 pc. Gilmozzi et al (1994) compiled 
various sources of maximum magnitude-rate of decline relationships and 
$t_{3}$ obtaining a distance of 680 pc.  We have summarized the various 
estimates of $t_{3}$, speed class and distances in Table 4.

What do we make of all of this? Certainly the correct
empirically-determined distance to V841 Oph remains unclear. While old
novae tend to be much more distant than what we have derived for it, there
are old novae which are comparably as close, those being V603 Aql (238 pc
from a Hipparcos parallax), DQ Her (327 pc), and HR Del (285 pc), the
latter two from Warner (1987).  We are encouraged that our model-derived
distance is close to the distance range estimated by Sherrington and
Jameson (1983). Our distance for V841 Oph is also reinforced by the HST
STIS study of the old nova DI Lac (Moyer et al. 2003) which is almost a
spectroscopic twin of V841 Oph. While the accretion rate is higher in DI
Lac (Nova Lac 1910) by a factor of 10 to 30 compared with V841 Oph (Moyer
et al. 2003), the distance implied by the scale factor of the best-fitting
accretion disk models in that study (after correcting a normalization
error) are in the range of 199 to 292 pc. If our model-derived distance
is incorrect, then there may be something wrong with using steady state
accretion disk models for old novae like V841 Oph and DI Lac or there are
missing pieces of physics in the accretion disk models we have used.

\section{Summary}

The two spectra of V841 Oph, taken 15 years apart, reveal a difference in
flux level with the later one having a higher flux level but no evident P
Cygni profile structure. The latter could be due to orbital
phase-dependent changes in the profile structure rather than a shutdown of
the outflow. Our synthetic spectral analysis of V841 Oph with model
accretion disks and photospheres reveals that the FUV spectrum is
completely dominated by an optically thick, steady state disk. This is no
surprise. However, since previous observational evidence supports a low
inclination and high intrinsic luminosity for the system, it is surprising
that the best-fitting steady state accretion disk implies an accretion
rate of $\sim$3 $\times$ 10$^{-11}$ M$_{\sun}$/yr. This is at least two
orders of magnitude lower than the accretion rate one typically associates
with old novae based upon their average absolute magnitudes, and lower
than the typical accretion rates derived for nova-like variables such as
IX Vel and RW Sex. While the nova-like variables have had no recorded
outburst, the IUE spectra of V841 Oph and DI Lac were observed 132 years
and 76 years after their classical nova explosions. Therefore, if our
accretion rates are correct, then their post-nova accretion rates, roughly
a century after their nova episodes, have declined significantly to levels
below the rates associated with the high states of VY Scl-type nova-like
variables. ($>$ 1 kpc).

It is clear that further FUV spectroscopic observations and distance
determinations from parallaxes are needed to confirm the results suggested
by our archival study. 

\acknowledgements

A portion of this manuscript was prepared during a visit by one of us
(EMS) at the Institut d'Astrophysique, in Paris. We are grateful to
Michael Friedjung for useful discussions on the distances of old novae.
We thank an anonymous referee for a number of useful comments.
This research was supported by NASA ADP grant NNG04GE78G, in part by
NSF grant AST05-07514 and by summer
research funding from the NASA-Delaware Space Grant Consortium.


\clearpage

\begin{figure}
\epsscale{0.8}\plotone{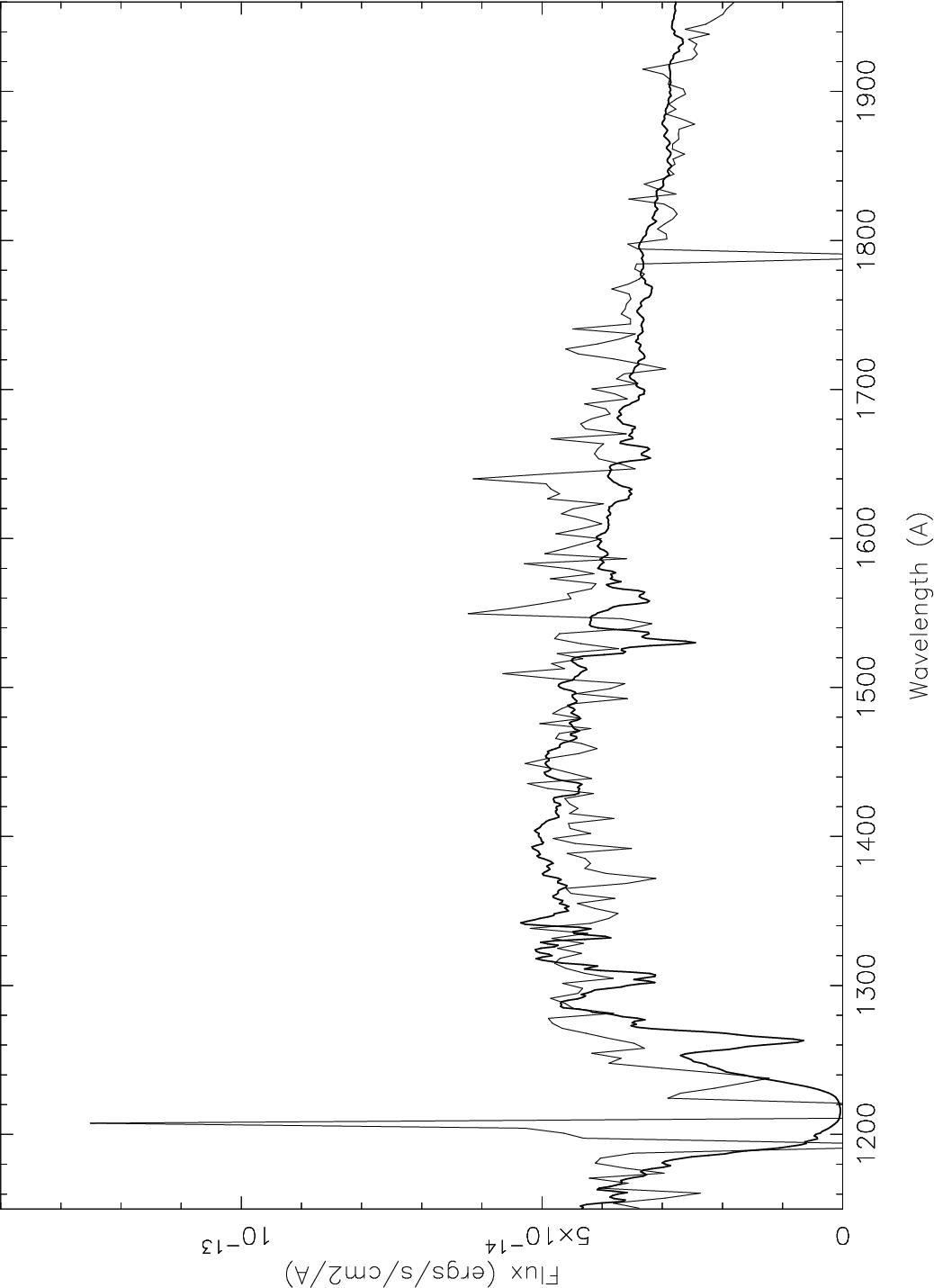}
\caption{Plot of Flux f$_{\lambda}$ (ergs/s/cm$^{2}$/\AA) versus
wavelength (\AA) for IUE spectrum SWP07950 shown with the best-fit
accretion disk model with $\dot{M} = 10^{-10.5}$ M$_{\sun}$/yr, $i =
18$\degr\, and M$_{wd} = 0.80$ M$_{\sun}$.}
\end{figure}     

\clearpage

\begin{figure}
\epsscale{0.8}\plotone{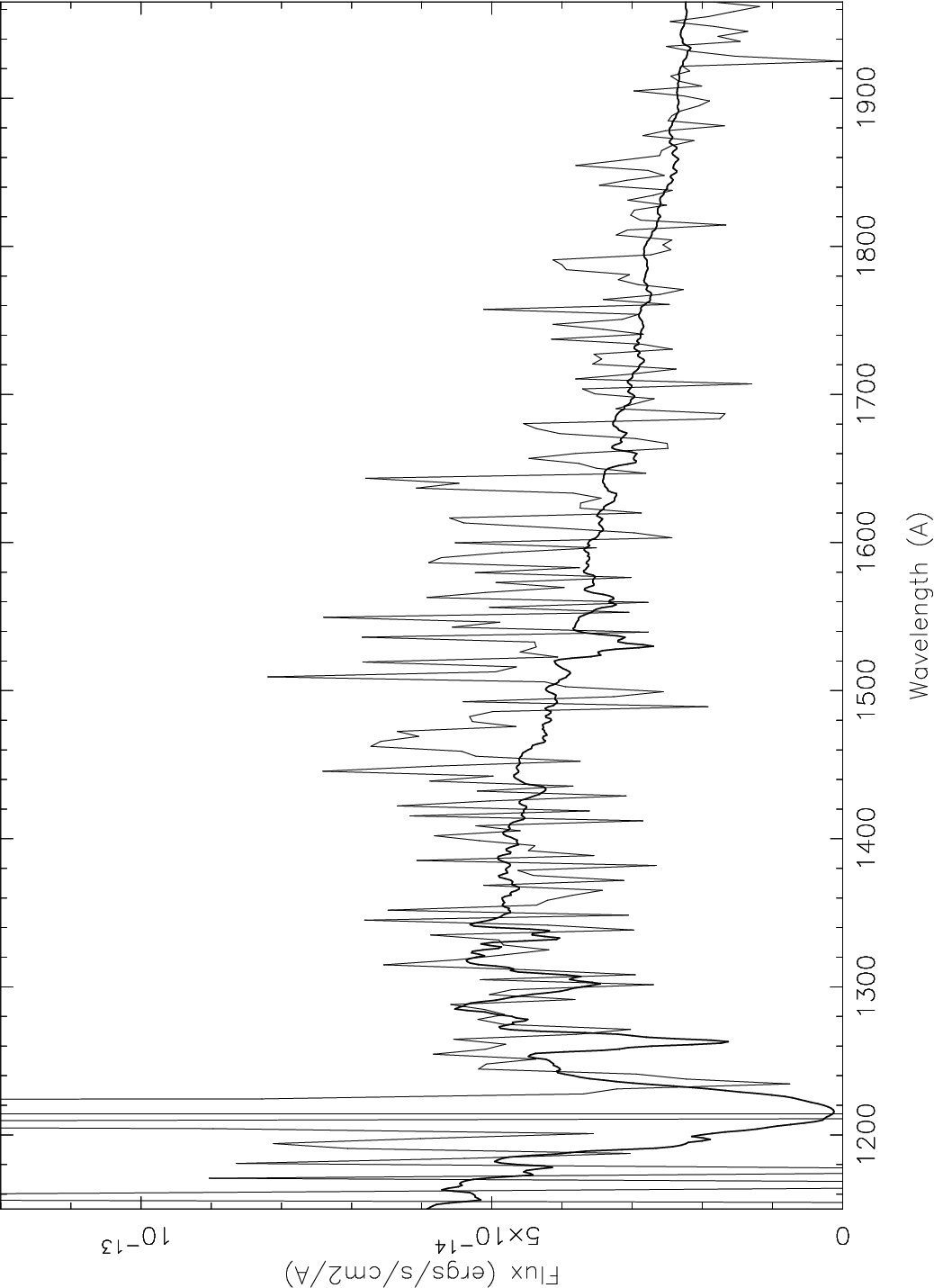}
\caption{Plot of Flux f$_{\lambda}$ (ergs/s/cm$^{2}$/\AA) versus
wavelength (\AA) for IUE spectrum SWP54454 shown with the best-fit
accretion disk model with $\dot{M} = 10^{-10.5}$ M$_{\sun}$/yr, $i =
18$\degr\, and M$_{wd} = 0.80$ M$_{\sun}$.}
\end{figure}      

\clearpage

\begin{figure}
\epsscale{0.8}\plotone{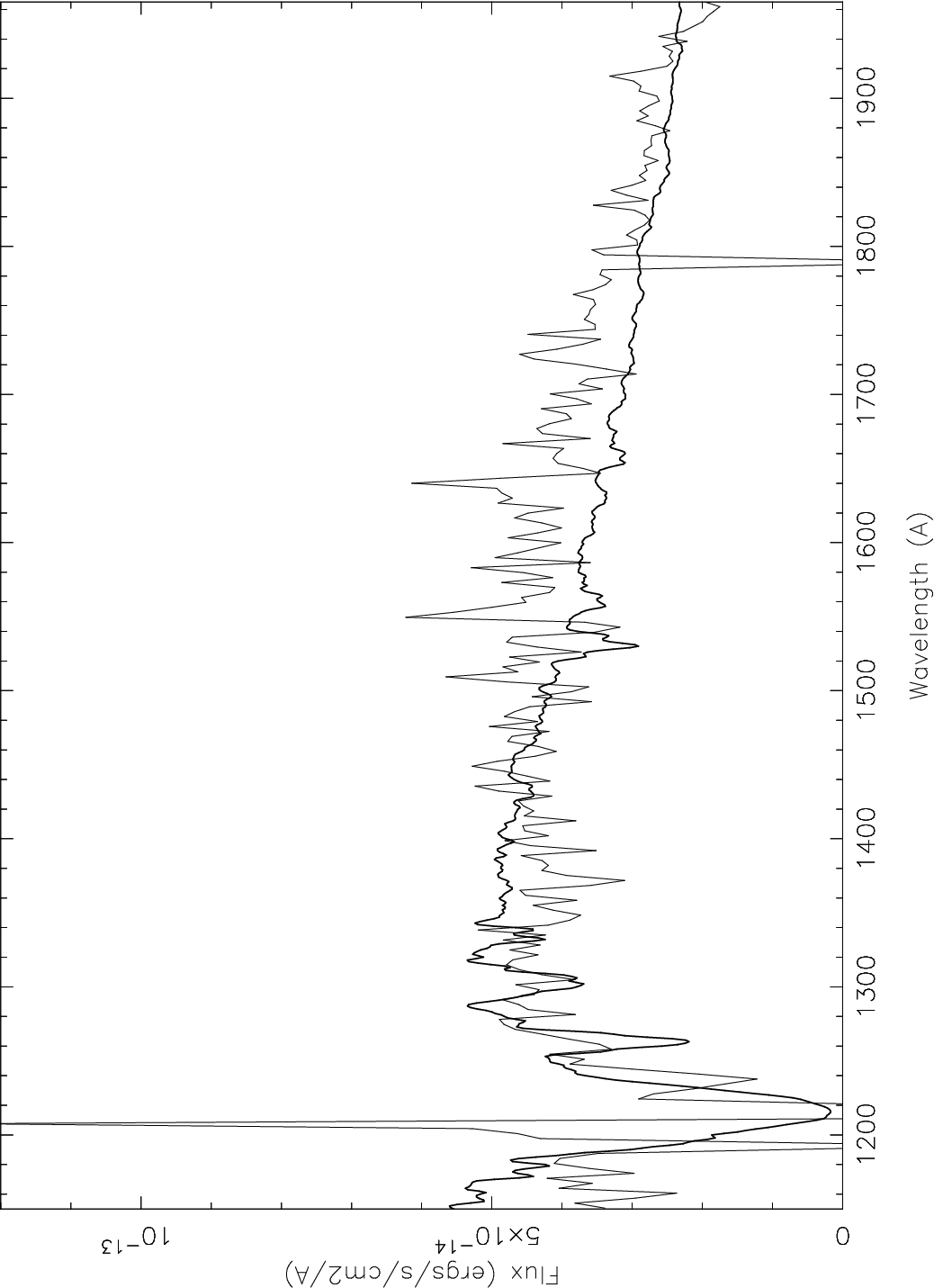}
\caption{Plot of Flux f$_{\lambda}$ (ergs/s/cm$^{2}$/\AA) versus
wavelength (\AA) for IUE spectrum SWP07950 shown with the best-fit
accretion disk model with $\dot{M} = 10^{-10.5}$ M$_{\sun}$/yr, $i =
18$\degr\, and M$_{wd} = 1.0$ M$_{\sun}$.}
\end{figure}        

\clearpage

\begin{figure}
\plotone{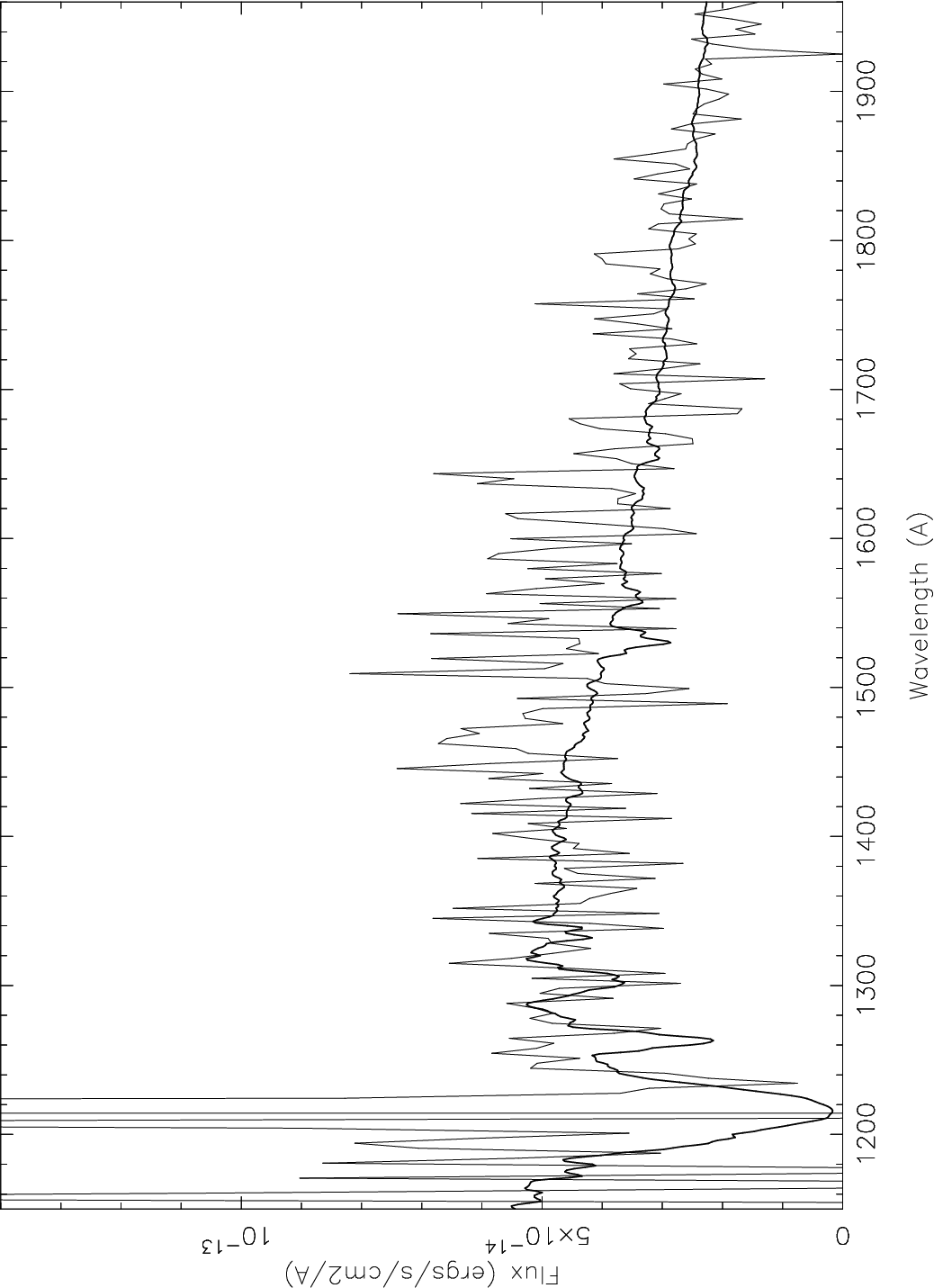}
\caption{Plot of Flux f$_{\lambda}$ (ergs/s/cm$^{2}$/\AA) versus
wavelength (\AA) for IUE spectrum SWP54454 shown with 
the best-fit accretion disk model with $\dot{M} =
10^{-10.5}$ M$_{\sun}$/yr, $i = 18$\degr\, and M$_{wd} = 1.0$ M$_{\sun}$.} 
\end{figure}    

\clearpage

\begin{figure}
\plotone{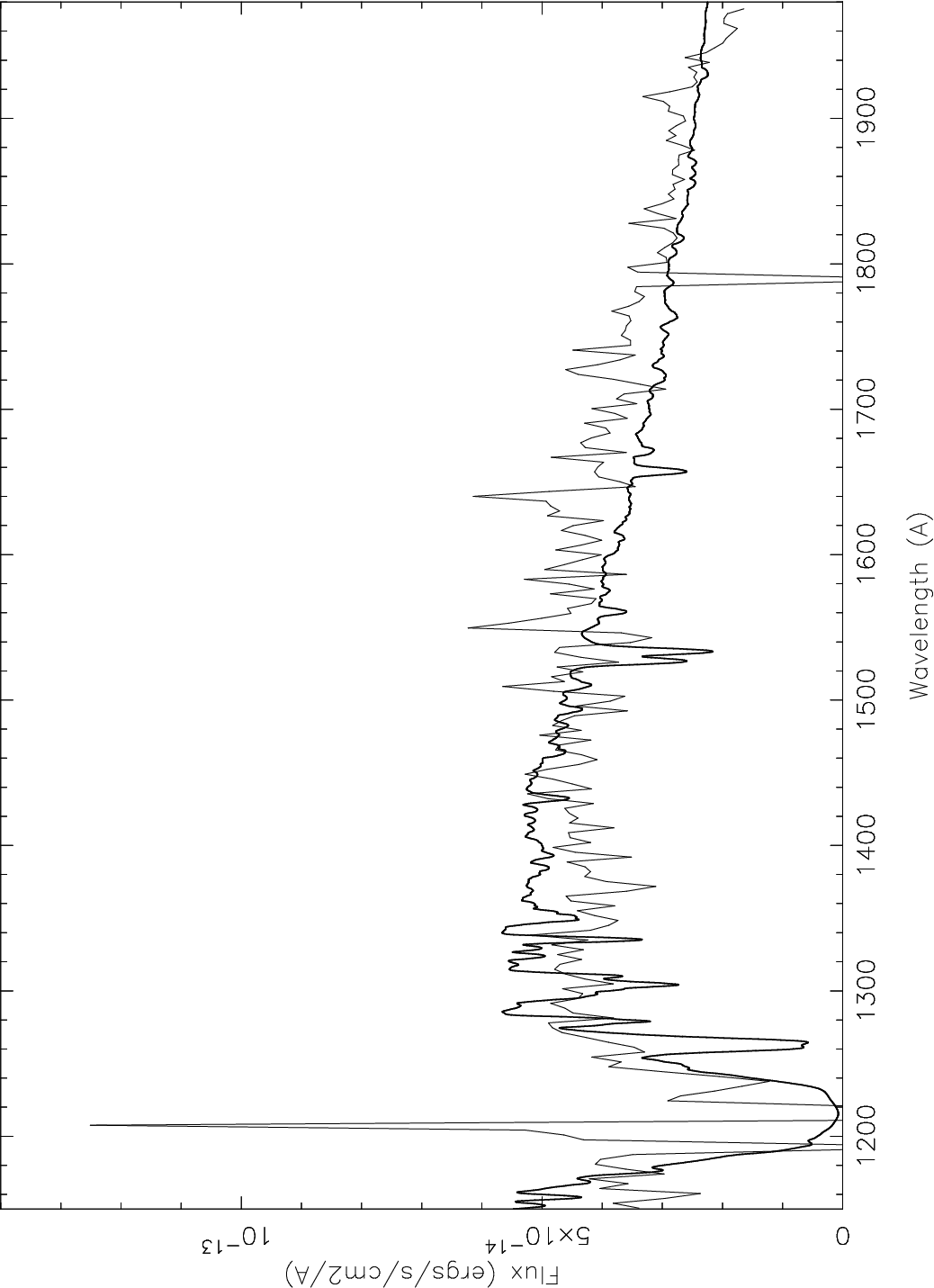}
\caption{Plot of Flux f$_{\lambda}$ (ergs/s/cm$^{2}$/\AA) versus
wavelength (\AA) for IUE spectrum SWP54454 shown with the best-fitting
high gravity (log $g = 8$) model with T$_{eff} = 18,000$ K.}
\end{figure}  

\clearpage

\begin{figure}
\plotone{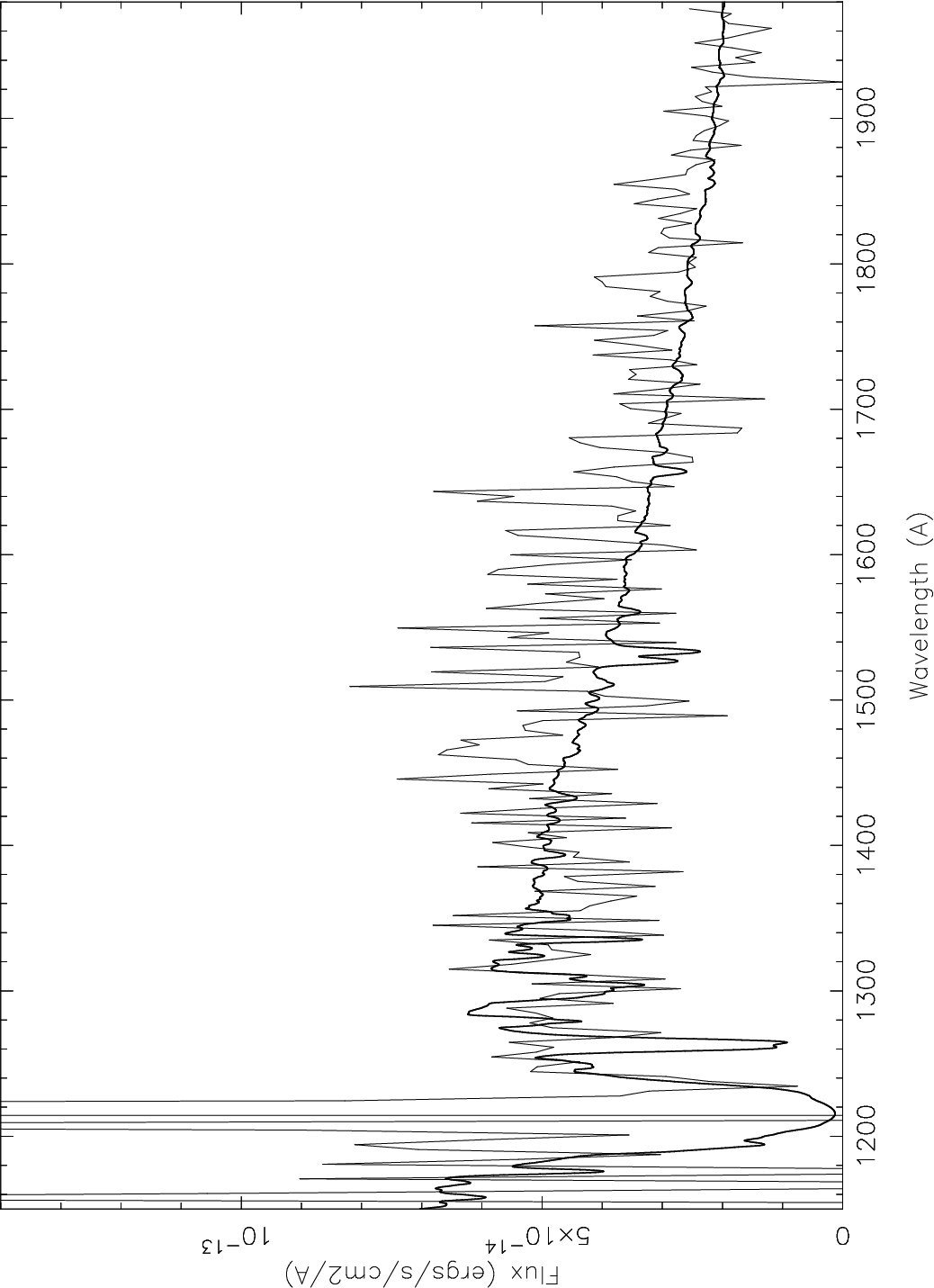}
\caption{Plot of Flux f$_{\lambda}$ (ergs/s/cm$^{2}$/\AA) versus
wavelength (\AA) for IUE spectrum SWP07950 shown with the best-fitting
high gravity (log $g = 8$) model with T$_{eff} = 20,000$ K.}
\end{figure}

\clearpage

\begin{deluxetable}{cc}
\tablewidth{0pc}
\tablecaption{\sc Parameters of V841 Oph}
\startdata
\hline
system                  &               V841 Oph\\
log P [{\it days}]      &               -0.219\tablenotemark{a}\\
{\it i}[\degr]          &               $\sim$0\tablenotemark{b}\\
M$_V$*                  &               3.9\\
M$_{WD}$                &               1?\\
log $\dot{M}$           &               -8.03\\
\enddata
\tablerefs{
(a) \citealt{Hoard2002};(b) \citet{Warner1995}\\
** Corrected for inclination effects}
\end{deluxetable}

\begin{deluxetable}{ccccccc}
\tablecaption{\sc IUE Observing Log}
\tablehead{
\colhead{Data}&\colhead{t$_{exp}$}&\colhead{Dispersion}&\colhead{Aperture}&\colhead{Date of Exposure}&\colhead{Cont. Cts}&\colhead{Back. Cts}}
\startdata
SWP54454        &       3600    &       LOW     &       LARGE   &       1995-04-17 19:41:00     &       122     &       77\\
SWP07950        &       7200    &       LOW     &       LARGE   &       1980-02-14 20:54:00     &       92      &       34\\
\enddata
\end{deluxetable}

\clearpage

\begin{deluxetable}{ccc}
\tablewidth{0pc}
\tablecaption{\sc Accretion Disk Fitting Parameters}
\tablehead{\colhead{Parameter}&\colhead{SWP54454}&\colhead{SWP07950}}
\startdata

M$_{WD}$(M$_{\sun}$)    &       0.8     &       1.0\\
{\it i}\degr\   &       18      &       18\\
$\dot{M}$(M$_{\sun}$/yr)        &       -10.5   &       -10.5\\
$\chi^2$&       1.49179 &       2.48270\\
S       &       0.712911        &       0.266539\\
{\it d} &       188     &       194\\
\enddata
\end{deluxetable}

\begin{deluxetable}{ccc}
\tablewidth{0pc}
\tablecaption{\sc V841 Oph Distance Estimates}
\tablehead{\colhead{Source}&\colhead{$t_{3}$ ({\emph d})}
          &\colhead{{\emph d}(\emph pc)}}
\startdata

Downes and Duerbeck (2000)       &  $\sim$12  &  380 (Type B Light Curve)\\
Fast Nova $t_{3}$ Relationship   &  $\sim$12  &  1000                    \\
Sherrington and Jameson (1983)   &  --        &  301--336                \\
Slow Nova $t_{3}$ Relationship   &  23        &  $>$370                  \\
Duerbeck (1981)                  &  130       &  790                     \\
Warner (1987)                    &  --        &  710                     \\
Gilmozzi et al (1994)            &  --        &  680                     \\
\enddata
\end{deluxetable}

\clearpage

\end{document}